\documentclass[twocolumn,showpacs,pre]{revtex4}

\usepackage{amsmath}
\usepackage{amssymb} 
\usepackage{amsfonts}
\usepackage{graphicx} 
\usepackage[hyperindex,breaklinks]{hyperref} 
\usepackage{theorem}
                 \def\d{\delta}     
\def\e{\varepsilon}            \def\h{\eta}     
        \def\l{\lambda}                  
\def\n{\nu}           \def\p{\pi}           
\def\s{\sigma}                        
\def\c{\chi}          \def\ps{\psi}              
                    
\def\D{\Delta}        \def\L{\Lambda}       
                           
\def\th{\vartheta}            
                  \def\o{\omega}

\def\LL{{\cal L}}

\def\bfe{{\bf e}} 
\def\pp{{\bf p}}\def\qq{{\bf q}}\def\xx{{\bf x}}
 
\def\yy{{\bf y}}\def\kk{{\bf k}}
\def\zz{{\bf z}}\def\uu{{\bf u}}\def\vv{{\bf v}}\def\ww{{\bf w}}

       \def\uo{{\underline \omega}}

          \def\uo{{\underline \o}}

\def\hw{{\widehat w}}

\def\hp{{\widehat \ps}}

\def\be{\begin{equation}}    \def\ee{\end{equation}}
\def\bea{\begin{eqnarray}}   \def\eea{\end{eqnarray}}
\def\bean{\begin{eqnarray*}} \def\eean{\end{eqnarray*}}
\def\bfr{\begin{flushright}} \def\efr{\end{flushright}}
\def\bc{\begin{center}}      \def\ec{\end{center}}
\def\bal{\begin{align}}      \def\eal{\end{align}}
\def\ba#1{\begin{array}{#1}} \def\ea{\end{array}}
\def\bd{\begin{description}} \def\ed{\end{description}}
\def\bv{\begin{verbatim}}    \def\ev{\end{verbatim}}

\def\lft{\left}                  \def\rgt{\right}
\def\la{{\langle}}               \def\ra{{\rangle}}

\def\Halmos{\hfill\vrule height10pt width4pt depth2pt \par\hbox to \hsize{}}
\def\pref#1{(\ref{#1})}
\def\lb#1{\label{#1}}

\def\qed{\raise1pt\hbox{\vrule height5pt width5pt depth0pt}}
\let\dpr=\partial

\let\==\equiv

\let\0=\noindent

\def\*{{\hfill\break\null\hfill\break}}

\def\insertplot#1#2#3#4#5#6{%
\begin{figure}[ht]
\begin{center}
\includegraphics[bb=0 0 #1 #2, clip,  
width=#1pt, height=#2pt,
  clip=true, viewport=0 0 #1 #2]{#4.ps}
  \caption{#5}
  \end{center}
  \end{figure}
  }

\begin{document} 
  \title{Correlation Critical Exponents for the Six-Vertex Model}
  %\title{A new theoretical approach to Interacting Dimers}
  %\title{Interacting Fermions Picture for Dimer Models}%

  \author{P. Falco 
  }
  \affiliation{Department of Mathematics, 
  California State University, Northridge,
  CA 91330
  }

  \begin{abstract}
The six-vertex model on a square lattice is ``exactly solvable'' because
an exact formula for the free energy can be obtained by Bethe Ansatz. 
However, exact formulas for the correlations of local bulk observables, 
such as the orientation of the arrow at a given edge, are in general 
not available.
In this paper we consider the isotropic ``zero-field'' six-vertex model at 
small $|\D|$. We derive the  large-distance asymptotic formula of 
the arrow-arrow correlation, which displays a  power law decay and 
an anomalous exponent. 
Our method is based 
on an interacting fermions representation of the six-vertex model and does not  
use any information obtained from  the exact solution.
  \end{abstract}
  \pacs{05.50.+q, 71.27.+a, 64.60.F-, 64.60.Cn}

  \maketitle

  \section{Introduction}
The six-vertex model on a square lattice is 
called ``exactly solvable'' because of the 
discovery in 1967 of the
exact formula for the free energy 
\citep{Lie67e,Lie67c,Lie67,Sut67,Lie67b,Yan67,SYY67}. 
Since then, there has been an intense research activity on 
this model and its transfer matrix \cite{LW72,Bax82,Res10}. 
That led to important results also for  systems 
that are equivalent to  the six-vertex model 
in an exact or approximate form \cite{LW72,Bax82,dN83}.

Many  properties of the six-vertex model are determined  by a characteristic 
parameter, $\D$. It is of special interest the {\it critical case} $|\D|\le 1$,
in which the  spectrum of the transfer matrix was found to be gapless, and
correlations 
of local bulk variables are therefore 
expected to display a large-distance power law decay. 
However,
a direct study of correlations that validates this prediction 
has been in general very problematic. 

In the simpler $\D=0$ case, in which  the six-vertex 
model is a {\it free}  fermion system,  
Sutherland \cite{Sut68}
derived an exact formula for 
the arrow-arrow correlation which displayed staggered prefactors and power law decay.  
In the  general $\D\neq 0$ case, 
except for the case of some non-local variables \citep{CP12}
and except for the non-critical regime $\D<-1$ \citep{JM95},
exact formulas for correlations  are 
not known.  
Nonetheless, large-distance properties of the 
arrow-arrow correlations
have been  inferred \citep{YAMC80} 
on the basis of two equivalences: a)  the  transfer matrices of 
the six-vertex model and of the Heisenberg quantum chain commute
\cite{MCW68}, hence the correlation of  arrows placed 
along the same lattice line 
must coincide with the $S^z$-spin correlation of the Heisenberg model;
b) the Heisenberg quantum chain is in the {\it Luttinger Liquid}  universality class 
\citep{LuPe75,Fog78,Hal80,BFM10}
(or can be studied via  conformal field theory \cite{BIR87})
and that  fixes the power law decay of correlations.

More recently,  a different method has been introduced to describe the 
critical phases of the six-vertex model,  
the Coulomb Gas picture for the {\it height variable} \citep{Nie84}. 
However, it appears that no specific result that can be compared with 
\citep{YAMC80}'s analysis has been derived along this route. 

In this paper we consider the isotropic zero-field six-vertex model
and derive large distance asymptotic formulas for 
 arrow-arrow correlations at small $|\D|$. 
Remarkably, our formulas are compatible with 
the features predicted in \citep{YAMC80}, including the  occurrence 
of an  {\it anomalous} critical exponent.
Our approach is made of three steps. 
First we recast the model into an {\it interacting} dimers model following \cite{Bax72}; 
then we transform the dimer problem into a interacting fermions systems; 
finally, we study fermion correlations by 
the standard  Renormalization Group techniques used in  condensed matter 
theory \cite{So79,MDC93,Gia04}. 
Although this way of dealing with the six-vertex model was already suggested 
in \cite{Fa13}, we observe  
that the technical implementation of the idea is new:   
the dimer interaction obtained from the six-vertex model 
equivalence turns out to be staggered, 
hence  a different fermion representation, with  different symmetries,  
is used. We also stress that our approach does not  
need any result from the exact solution.   

  \section{Definitions and Results }
Consider a finite square lattice $\L'$. 
A configuration $\o$ of the six-vertex model is obtained 
by drawing an arrow per each edge of the lattice so that the number of 
incoming arrow at each vertex is $2$ (the {\it ice-rule}). 
At  each vertex there is one of the six possible
arrangements of arrows in Fig. \ref{f3}: assign them   
positive weights $p_1, p_2, \ldots, p_6$.
\insertplot{240}{135}
{}{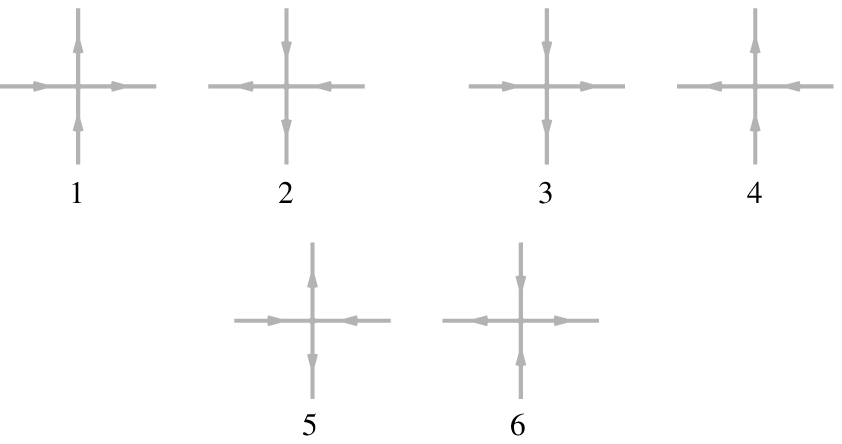}{The six-possible configurations of arrows.\lb{f3}}{}
If $n_j(\o)$ is the number of vertex in the configuration $\o$ that  
display the arrangement number $j$,  the partition function of the six vertex model is 
$$
Z=\sum_{\o} \prod_{j=1}^6 p_j^{n_j(\o)}\;.
$$ 
We assume periodic boundary condition, so that 
$n_5(\o)=n_6(\o)$ and  $p_5p_6$ counts as one free parameter. 
Our results --as well as the analysis in \citep{YAMC80}-- 
are  for the {\it zero-field six-vertex model}, i.e. the case
$$
p_1=p_2=a\qquad p_3=p_4=b\;;
$$
furthermore, for sake of simplicity,  in this paper 
we will only consider the {\it isotropic case} $a=b$.
Without loss of generality,
we can fix $p_5=a$  
and parametrize $p_6=2a e^\l$,  for a real $\l$.    
The characteristic parameter of the six vertex model is 
$$
\D=\frac{a^2 + b^2 - p_5 p_6}{2ab}=1-e^{\l}\;.
$$
The case $\D=0$  is the free fermion case.
Our goal is to show that,
at least for small  $|\D|$,  
there are correlations 
with power law decay. 
We consider the {\it arrow orientation} observable. 
Let $\vv_0$ and $\vv_1$ be the two orthogonal vectors that span $\L'$; 
a point of the plane will be parametrized by 
$\xx=x'_0 \vv_0 + x'_1 \vv_1$. 
For $d$ a horizontal edge centered at a point $\xx$,  
the horizontal arrow orientation $\s_0(\xx)$ is equal to $1$ if the arrow along $d$
points to the right, otherwise it is equal to   $-1$; similarly, 
for $d$ a vertical edge centered at a point $\xx$, the vertical  
arrow orientation $\s_1(\xx)$ is equal to  $1$ if the arrow along  $d$
points up, otherwise it is equal to $-1$.
In the infinite lattice limit,  the arrow correlations have 
the following large $|\xx|$ asymptotic 
formula: for two vertical arrows
\begin{align}\lb{ddc}
\la \s_1(\xx+\yy)\s_1(\yy)\ra\sim 
&c_0 \frac{{x'}_1^2-{x'}_0^2}{({x'}_0^2+{x'}_1^2)^2}
+c_- 
\frac{(-1)^{x'_0+x'_1} }{({x'}_0^2+{x'}_1^2)^{\kappa_-}}\;,
 \end{align}
for one horizontal and one vertical arrow
  \begin{align}\lb{ddc2}
  \la \s_1(\xx+\yy)&\s_0(\yy)\ra 
%-\la \s_1(\xx+\yy)\ra \la \s_0(\yy)\ra
%  \notag\\
%  &
\sim c_0 
  \frac{-2 x'_0 x'_1}{({x'_0}^2+{x'_1}^2)^2}-c_- 
  \frac{(-1)^{x'_0-x'_1} }{({x'_0}^2+{x'_1}^2)^{\kappa_-}}\;.
  \end{align}
Each of the two above formulas is made of two terms: 
the former term has the same power law decay of the $\D=0$ case; the 
latter term has a power law decay  with an {\it anomalous} (i.e.  $\D$-dependent) 
 critical exponent $\kappa_-=1+O(\D)$
and  a staggering prefactor $(-1)^{x'_0+x'_1}$ or $(-1)^{x'_0-x'_1}$; besides, 
$c_0$ and $c_-$ are $\frac2{\p^2}+O(\D)$. Note that $x'_0+x'_1$ and $x'_0-x'_1$ are integers.
As by-product of our approach, we can obtain a Feynman
graphs representation of the expansion of $\kappa_-$ in powers of
$\l$. For example, at first order
\be\lb{ddc22}
\kappa_-=1-\frac{2\l}{\p}+O(\l^2)\;;
\ee
therefore, if $\l$ is positive and small (i.e. $\D$ negative and small), then  
at large distances 
the latter terms  in \pref{ddc} and \pref{ddc2} dominate 
over the former ones.  
Formulas \pref{ddc}, \pref{ddc2} and \pref{ddc22} are the main result of this paper. 
For $\D=0$, \pref{ddc} coincides with Sutherland's exact solution,  
see (6) of \cite{Sut68}.   For $\D\neq0$, \pref{ddc}, \pref{ddc2} and \pref{ddc22}
as expected from  \citep{YAMC80}, 
are  in   agreement with the  
asymptotic formulas for the  
$S^z$--spin correlations 
in the Heisenberg quantum chain \citep{LuPe75,Fog78,Hal80,BFM10}. 

As an application of this result, we consider the 
large distance behavior of the covariance of the {\it height variable} 
\cite{vB77}. 
For $\uu$ the center of a plaquette of $\L'$, $h(\uu)$ is the 
integer variable such that: if $\xx$ is the center of a vertical bond 
$$
\vv_0\cdot \nabla h(\xx)\=
h(\xx+\frac12\vv_0)-h(\xx-\frac12\vv_0)=\s_1(\xx)\;;
$$
if $\xx$ is the center of a horizontal bond
$$
-\vv_1\cdot \nabla h(\xx)\=
h(\xx-\frac12\vv_1)-h(\xx+\frac12\vv_1)=\s_0(\xx)\;.
$$
Because of the ice rule,  $h(\uu)$ is a scalar potential 
defined up to a global constant. From \pref{ddc}
we find the large $|\xx|$ asymptotic formula 
\be\lb{ddc35} 
\la \big[h(\xx+\uu)-h(\uu)\big]^2\ra\sim 2c_0 \ln |\xx|\;;
\ee
hence  $h(\uu)$ has the same large distance behavior of a 
{\it free} boson field 
in dimension two. It is remarkable that, 
because of the staggering prefactor, the 
latter term in \pref{ddc} does not determine the leading term of 
\pref{ddc35} regardless of the sign of $\l$. 

In the next sections we will derive \pref{ddc},  \pref{ddc2}, \pref{ddc22};
and we will show how to obtain the application  \pref{ddc35}. 
\section{Interacting Fermions Picture}
Our 
point of departure is the equivalence  of the six-vertex model 
on the square lattice $\L'$ with 
an interacting dimers model (IDM) on a different square lattice $\L$ \citep{Bax72}. 
A dimer configuration $\o$ on $\L$ is a collection of dimers covering
some of the edges of $\L$ with the constraint that every vertex of
$\L$ is covered by one, and only one, dimer. The general partition
function of the IDM is 
\be\lb{allen}
Z_\l=\sum_{\o} \exp\Big\{\l \sum_{d,d'\in \o} v(d,d')\Big\}
\ee
where: the first sum is over all the
dimer configurations; the second sum is over any pair of
dimers in the configuration $\o$; finally,   
$\l$ is the dimers coupling constant and  $v(d, d )$ is an 
interaction that we have to determine 
to have the equivalence with the six-vertex model. 
To do so, first 
embed $\L'$ into $\L$ 
as showed in Fig. \ref{f2}; 
\insertplot{220}{200}
{}{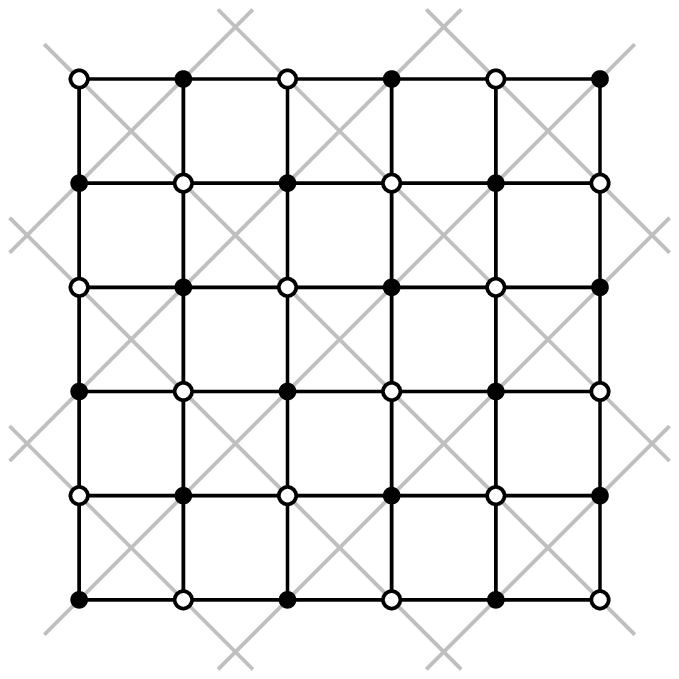}{Superposition of the six-vertex lattice (in gray) and the interacting dimers lattice 
(in black). Note the role played by the bi-partition of the latter. \lb{f2}}{0} 
then use the mapping from the six-vertex configurations to the 
dimer configurations in Fig. \ref{f5}. 
\insertplot{340}{160}
{}{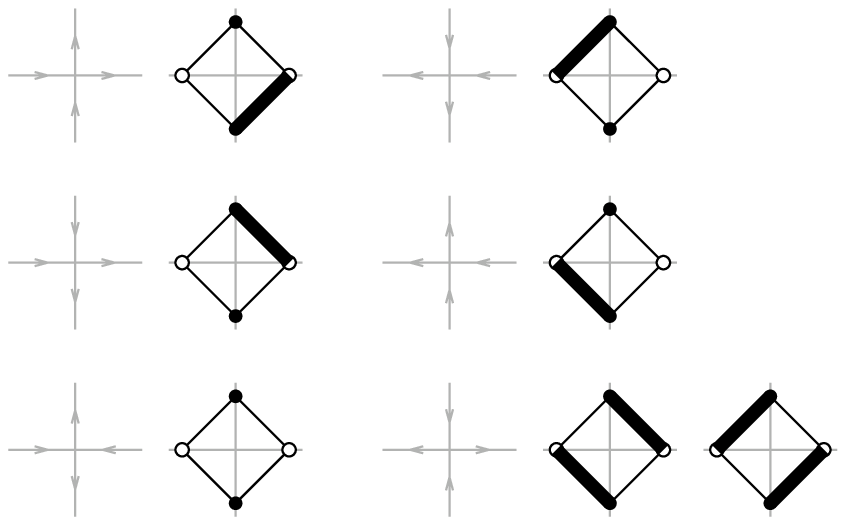}{Map of the six-vertex configurations on $\L'$ 
into the dimer model configurations on $\L$. The vertex configurations number $j=1,\ldots,5$ 
correspond to dimer arrangements with weight 1; the vertex configuration number 
6 corresponds to two dimers arrangement, each of which with a weight $e^\l$. 
Note that in this Figure 
the black sites are in vertical position; when the black sites are in 
horizontal position all the dimer arrangements have to be counted with a weight 1.  
 \lb{f5}}{0} 
As a consequence set $v(d,d')=1$ if $d$ and $d'$ are the two parallel nearest-neighbor 
dimers in the last two arrangement in Fig \ref{f5}, which have  the black
sites in vertical positions;  and  $v(d,d')=0$ otherwise.
Let $\bfe_0$ and $\bfe_1$ the two vectors that span $\L$; a point in the plane 
will be $\xx=x_0\bfe_0 + x_1\bfe_1$;  in particular $\vv_0=\bfe_0-\bfe_1$ 
and $\vv_1=\bfe_0+\bfe_1$. 
A natural local bulk observable for the dimer model is the {\it dimer occupancy} 
$\n_j(\xx)$ which is equal to $1$ if the edge $\{\xx,\xx+\bfe_j\}$ is occupied, 
and 0 otherwise. The relationships between the dimer occupancy
and the arrow orientation of the equivalent six-vertex model is: 
for $\xx$ a white site, 
\begin{align}\lb{arkani1}
\s_0(\xx):=\n_0(\xx-\bfe_0)+\n_1(\xx)-\n_0(\xx)-\n_1(\xx-\bfe_1)
\end{align}
for $\xx$ a black site, 
\begin{align}\lb{arkani2}
\s_1(\xx):=\n_0(\xx)+\n_1(\xx)-\n_0(\xx-\bfe_0)-\n_1(\xx-\bfe_1)\;.
\end{align}
(As a side remark, also for the dimer model 
one can introduce an integer height function $H(\ww)$
for every $\ww$ that is the  center of a plaquette of the lattice $\L$. 
The standard definition is such that, 
if $\uu$ is the center of a plaquette of $\L'$ {\it and}
the center of a plaquette of $\L$, one has 
$$
H(\uu)=2 h(\uu)\;.
$$
We will not need $H(\ww)$ in the rest of the paper.)   
Since black and white lattice sites play a different role in the choice of the 
dimer potential, 
we need a fermion representation of the dimer model that takes into account 
the bi-partition of the square lattice.  For this reason the fermion representation 
that we introduce below is different from the one used in \citep{Fa13}.

Let $\LL$ be the Bravais lattice of the black sites of $\L$.  $\LL$ 
is spanned by the vectors $\vv_0$  
and $\vv_1$;  its primitive cell has volume $v_P=2$;
and  
the reciprocal lattice of $\LL$ is spanned by  $\hat\vv_0=\frac12\vv_0$ 
and $\hat\vv_1=\frac12\vv_1$.

When $\l=0$ the dimer model is equivalent to a lattice fermion 
field without interaction. 
Namely
\be\lb{pf0}
Z_{0}=\int\! D\ps\; \exp\Big\{-\sum_{\xx,\yy\in \LL} K_{\xx, \yy} \ps^+_\xx \ps^-_{\yy+\bfe_0}\Big\}
%=\Pfa_{\xx, \yy} K_{\xx, \yy}
\ee
where: $\{\ps^+_\xx, \ps^-_{\xx+\bfe_0}:\xx\in \LL\}$ 
are Grassmann variables and  $D\ps$ indicates the 
integration with respect to all of them; 
$K_{\xx, \yy}$ is one of the possible 
{\it  Kasteleyn matrix} for the square lattice 
dimer model  
$$
K_{\xx, \yy}=\d_{\xx,\yy}-\d_{\xx-2\bfe_0,\yy}- \d_{\xx-\bfe_0+\bfe_1,\yy}- \d_{\xx-\bfe_0-\bfe_1,\yy}\;.
$$
\pref{pf0} is  the partition function of
a free {\it fermion field}, i.e. a Grassmann-valued 
Gaussian field  with moment 
generator 
$$
\la e^{i\sum_{\xx}(\ps^+_\xx\h^-_\xx+\h^+_\xx\ps^-_{\xx+\bfe_0})}\ra_{0} = 
e^{\sum_{\xx,\yy}S(\xx-\yy)\h^+_\xx\h^-_\yy}
$$  
where: the $\h_\xx$'s are external  Grassmann variables; 
$S$ is the inverse Kasteleyn 
matrix  
$$
S(\xx)=\frac{v_P}{2}\int_{\rm 1BZ}\!\frac{d\kk}{(2\p)^2}\; 
\frac{e^{i \kk\cdot (\xx+ \bfe_0)}}
{i\sin(\kk\cdot \bfe_0)-\cos (\kk\cdot \bfe_1)}
$$
and 1BZ is the  first 
Brillouin zone.
The Fourier transform 
of $S$ has two poles at the {\it Fermi momenta} 
$\kk=\o \pp_F$  for $\o=\pm1$ and $\pp_F=(0,\frac\p2)$.
Therefore, in view of the study of the scaling limit, 
it is convenient  to decompose
$$
S(\xx)=\sum_{\o=\pm} e^{i \o \pp_F\cdot \xx} S_\o(\xx)
$$
for
$$
S_\o(\xx)=\frac{v_P}{2}\int_{\rm 1BZ}\!\frac{d \kk}{(2\p)^2}\; 
\frac{e^{i \kk\cdot (\xx+ \bfe_0)} \c(\kk)}
{i\sin(\kk\cdot \bfe_0)-\o\sin (\kk\cdot \bfe_1)}
$$ 
where $\c(\kk)$ is 1 in a neighborhood of $\kk=0$ and 
such that $\c(\kk-\pp_F)+\c(\kk+\pp_F)=1$. Correspondingly, 
for $\e=\pm$, 
\begin{align}
\ps^\e_{\xx}&=\sum_{\o} e^{i\e \o \pp_F\cdot \xx} \ps_{\xx,\o}^\e
\end{align}
where $(\ps^+_{\xx,+},\ps^+_{\xx,-} )$ and 
$(\ps^-_{\xx,+},\ps^-_{\xx,-} )^T$
are   Dirac spinors 
with covariances
\begin{align}
&\la \ps_{\xx,\o}^+ \ps_{\yy,\o'}^+\ra_0
=\la \ps_{\xx+\bfe_0,\o}^- \ps_{\yy+\bfe_0,\o'}^-\ra_0=0 
\notag\\
&\la \ps_{\xx,\o}^+ \ps_{\yy+\bfe_0,\o'}^-\ra_0= \d_{\o,\o'} S_\o(\xx-\yy)\;.
\end{align}
If we now let $\l\neq 0$
one can verify that \pref{allen} becomes
\be\lb{pf1}
Z_{\l}=Z_0\la 
e^{(1-e^\l)V(\ps)}
\ra_0
\ee 
where $V(\ps)$ is quartic in the Grassmann variables with a simple explicit 
formula which is given in the next section. 
Besides, it is not difficult to find the leading term 
for large distances of the 
the dimer correlations: if $T$ indicates a truncated correlation 
and $\zz=\xx-\yy$, for two horizontal dimers
\begin{align}\lb{ddc3}
&\la \n_0(\xx)\n_0(\yy)\ra - \la \n_0(\xx)\ra\la\n_0(\yy)\ra 
\notag\\
&\sim
(-1)^{z_0+z_1}
\sum_\o\la \ps^+_{\xx,\o} \ps^-_{\xx,\o}; \ps^+_{\yy,\o}\ps^-_{\yy, \o} \ra^T
\notag\\
&+
(-1)^{z_0}\sum_\o 
\la\ps_{\xx,\o}^+\ps^-_{\xx,-\o};\ps_{\yy,-\o}^+\ps^-_{\yy,\o}\ra^T\;,
\end{align}
whereas for one horizontal and one vertical dimer
\begin{align}\lb{ddc4}
&\la \n_1(\xx)\n_0(\yy)\ra - \la \n_1(\xx)\ra\la\n_0(\yy)\ra 
\notag\\
&\sim
(-1)^{z_0+z_1}
\sum_\o i\o\la \ps^+_{\xx,\o} \ps^-_{\xx,\o}; \ps^+_{\yy,\o}\ps^-_{\yy, \o} \ra^T
\notag\\
&+
(-1)^{z_0}\sum_\o i\o
\la\ps_{\xx,\o}^+\ps^-_{\xx,-\o};\ps_{\yy,-\o}^+\ps^-_{\yy,\o}\ra^T\;.
\end{align}
(The above formulas hold regardless of the colors of the sites $\xx$ and $\yy$.)
The Renormalization Group argument that we will provide in the next section 
will indicate that  to compute the correlations up to subleading terms, 
we can just replace the fields $\ps_{\xx,\o}^+$, $\ps_{\xx,\o}^-$ with the 
Thirring model fields  $\ps_{\o}^\dagger(\xx)$, $\ps_{\o}(\xx)$ times a prefactor 
$2^{-\frac 14} e^{i\o \frac\p8}$ per each of them;  from the exact solution  
of the  Thirring model correlations 
\cite{Jo61,Kla64,Ha67,Fa06,BFM07,BFM09} (see also \cite{Fa12b})
\begin{align}
&\la \ps^\dagger_{\o}(0) \ps_{\o}(0); \ps^\dagger_{\o}(\xx)\ps_{\o}(\xx) \ra^T
=c_{T,0}
\frac{{x'_0}^2-{x'_1}^2-2i\o x'_0 x'_1}{({x'_0}^2+{x'_1}^2)^2}
\notag\\
%\la \ps^\dagger_{+}(0) \ps^\dagger_{-}(0); \ps_{-}(\xx)\ps_{+}(\xx) \ra^T
%&=
%\frac{c_+}{(x_0^2+x_1^2)^{\kappa_+}}
%\notag\\
&\la\ps_{\o}^\dagger(0)\ps_{-\o}(0);\ps_{-\o}^\dagger(\xx)\ps_{\o}(\xx)\ra^T
=
\frac{c_{T,-}}{({x'_0}^2+{x'_1}^2)^{\kappa_-}}
\end{align}
where the critical exponent $\kappa_-=1-\frac{\l_T}{2\p}+O(\l_T^2)$ 
and $\l_T$ is a parameter of the Thirring model: at first order 
$\l_T=4\l+O(\l^2)$ (see next section). 
\section{RG Analysis} 
The fermion interaction $V(\ps)$ has explicit formula
\begin{align}
&\sum_{\o_1, \o_2\atop \o'_1, \o'_2}
\frac{ v_P^4}{(2\p)^7}\int d\kk_1d\kk_2 d\qq_1d\qq_2\;
\hp_{\kk_1, \o_1}^{+}\hp_{\kk_2, \o_2}^{+}
\hp_{\qq_1, \o'_1}^{-}\hp_{\qq_2, \o'_2}^{-}
\notag\\
&\cdot
\d\lft(\sum_{j=1}^2 (\kk_j+\o_j\pp_F) -\sum_{j=1}^2 (\qq_j+\o'_j\pp_F)\rgt)
 v_{\uo;\uo'}(\underline\kk;\underline\qq)
\end{align}
where $v_{\uo;\uo'}(\underline\kk;\underline\qq)\=
v_{\o_1, \o_2, \o'_1, \o'_2}(\kk_1, \kk_2,\qq_1,\qq_2)$ is 
\begin{align}
&2\sin\lft[(\kk_1-\kk_2) \frac{\vv_1}2+(\o_1-\o_2)\frac\p4\rgt]
\notag\\
&\cdot\sin\lft[(\qq_1-\qq_2) \frac{\vv_0}2 -(\o'_1-\o'_2)\frac\p4\rgt]\;.
\end{align}
We follow  the RG method in 
\cite{Ga85}. Integrating out the large momentum scales, we obtain 
an effective interaction 
\begin{align}\lb{bois}
&\sum_{n\ge 1}
\sum_{\o_1\ldots, \o_{n}\atop \o'_1\ldots, \o'_{n}}
\int\frac{d\pp_1\cdots d\qq_{n}}{(2\p)^{4n-1}}
\; \hp_{\pp_1,\o_1}^{+}\cdots \hp_{\pp_n,\o_n}^{+}\hp_{\qq_{1},\o'_{1}}^{-}
\cdots \hp_{\qq_{n},\o'_{n}}^{-}
\notag\\
&\cdot \d\lft(\sum_{j=1}^n(\pp_j+\o_j\pp_F)- \sum_{j=1}^{n}(\qq_j+\o'_j\pp_F)\rgt) 
\hw_{n;\uo;\uo'}(\underline\pp; \underline\qq)  
\end{align}
where $\hw_{n;\uo;\uo'}$'s are series of Feynman graphs. 
Some symmetries are of crucial importance. 
For $R(k_0,k_1)=(k_1,-k_0)$,
and $\th (k_0,k_1)=(k_1,k_0)$, %and $\t (k_0,k_1)=(k_0,k_1+\p)$, 
because of the explicit formulas of $v_{\uo;\uo'}(\underline\kk;\underline\qq)$ 
 and 
of the Fourier transform of  $S_\o(\xx)$,
we have
\begin{align}\lb{sym}
%&\hw_{2m}(\t\kk_2, \ldots, \t\kk_{2m})=(-i)^m\hw_{2m}(\th\kk_2, \ldots, \th\kk_{2m})
%\notag\\
&\hw_{n;\uo;\uo'}(\underline{R \pp};\underline{R \qq})=e^{i\frac\p2 \sum_j\o_j}
\hw_{n;\uo';\uo}(\underline\qq;\underline\pp)
\notag\\
&\hw_{n;\uo;\uo'}(\underline{\th \pp};\underline{\th \qq})=e^{i\frac\p2 \sum_j\o_j}
\hw_{n;-\uo;-\uo'}(\underline\pp;\underline\qq)\;.
\end{align}
From power counting, 
there are two kinds of terms that are not irrelevant: the
quartic and quadratic ones.  
Using \pref{sym}, the quartic terms give a local contribution  
\begin{align}\lb{4}
\l'
\sum_{\xx,\o}\ps_{\xx,\o}^+\ps_{\xx,\o}^-\ps_{\xx,  -\o}^+\ps_{\xx, -\o}^- 
\end{align}
for
$\l'=\hw_{2;+,-;-,+}(0,0;0,0)-\hw_{2;+,-;+,-}(0,0;0,0)=-4\l +O(\l'^2)$
the effective coupling constant.
The quadratic terms cannot generate any local contribution of the 
form $\ps^+_{\xx,\o}\ps^{-}_{\xx,-\o}$ because of the delta function 
in \pref{bois} and the fact that the momenta are small; therefore,
using \pref{sym}, their only local contribution is  
\begin{align}\lb{2}
z\sum_{\xx,\o} 
\ps_{\xx, \o}^+ \dpr_{\o}\ps_{\xx, \o}^- \;,
\end{align}
for  
$z=\frac12\lft[-i\dpr_{p_0} \hw_{1;+;+}(\pp;\pp)
-\dpr_{p_1} \hw_{1;+;+}(\pp;\pp)\rgt]_{\pp=0}=O(\l^2)$ 
the field renormalization  counterterm 
and where $\dpr_\o$ is the  Fourier transform of $ik_0-\o k_1$.
An important fact to note is that we have not included in \pref{2}
any mass term: 
the localization of the quadratic terms would give 
$m\sum_{\xx,\o} 
\ps_{\xx, \o}^+ \ps_{\xx, \o}^- $, for $m=\hw_{1;\o;\o}(0;0)$; 
however, 
because of \pref{sym},
$\hw_{1;\o;\o}(\pp;\qq)= i\o\hw_{1;\o;\o}(R\qq;R\pp)$
and hence 
$$
m=0\;.
$$ 
At infrared scales, the beta function of this fermion model 
asymptotically coincides 
with the beta function of the Thirring model, which is vanishing \cite{MDC93}. Hence, 
scaling each $\ps^\e_{\xx,\o}$ of a factor $2^{-\frac 14} e^{i\o \frac\p8}$ so to match the
standard
normalization of the free part of the Thirring model,  by comparison with \cite{BFM07}, 
we obtain $\l_T=-\l'+O(\l^2)$, where the higher orders are determined by the irrelevant terms.
\section{Height Covariance}
For simplicity of notation we derive \pref{ddc35} in the case 
$\uu=(-N-\frac12)\vv_0+\frac12 \vv_1$ and 
$\xx=(2N+1) \vv_0$ only, but the formula for any $\uu$ and 
$\xx$ follows from the same argument. 
The height covariance is then given by 
$$
%\la \big[h(\xx+\uu)-h(\uu)\big]^2\ra=
\sum_{i,j=-N}^N
\la \vv_0\cdot \nabla h(\xx_i) \;\vv_0\cdot \nabla h(\xx_j)\ra\;,
$$
where $\xx_j=j\vv_0 + \frac 12 \vv_1$. 
In the double summation, the  $O(N)$ term cancels \cite{Spe13}: indeed since the summation 
of the height difference over any closed contour is vanishing by definition, and 
since the arrow correlation have a decay faster than the inverse distance,
one can replace  $j=-N, \ldots, N$ with $|j|\ge N+1$.  Hence
the height covariance becomes  
$$
%\la \big[h(\xx+\uu)-h(\uu)\big]^2\ra=
-\sum_{|i|\le N}\sum_{|j|\ge N+1}
\la \s_1(\xx_i) \; \s_1(\xx_j)\ra\;.
$$
Now plug \pref{ddc} into this formula: 
the staggered term gives a contribution that is bounded in $N$, 
whereas the un-staggered one gives $2c_0 \ln N$ plus terms that are bounded in $N$.
\section{Conclusion}
We have showed that the interacting fermions 
representation of the six-vertex model, 
in combination with 
a  Renormalization Group approach, 
provides a precise formula for the large distance decay 
of the arrow-arrow correlations 
for small $|\D|$. 
More in general, using ideas in \citep{Fa06,BFM07}, one could  show that 
the scaling limit of the $n$-points arrow correlations,  
apart from   staggering prefactors, 
are linear combinations of  Thirring model correlations.

\bibliographystyle{apsrev4-1}
\bibliography{svm.bib}

\end{document}